\documentclass[showpacs]{revtex4} 
\usepackage{epsfig,amsmath,amssymb,graphicx,upgreek,textcomp}
\usepackage{natbib}
\usepackage[english]{babel} %

\begin{document}

\vspace{0mm}
\title{Transition of a binary solution into an inhomogeneous phase} %
\author{Yu.M. Poluektov}
\email{yuripoluektov@kipt.kharkov.ua} %
\affiliation{National Science Center ``Kharkov Institute of Physics and Technology'', 61108 Kharkov, Ukraine} %
\author{A.A. Soroka} %
\affiliation{National Science Center ``Kharkov Institute of Physics and Technology'', 61108 Kharkov, Ukraine} %

\begin{abstract}
The thermodynamics of phase transitions of binary solutions into
spatially inhomogeneous one-dimensional states is studied
theoretically with taking into account nonlinear effects. %
It is shown that below the spinodal decomposition temperature there
first occurs  a phase stratification, and at a lower temperature,
after reaching the maximum possible stratification, there occurs a
second-order transition into the phase with concentration waves. The
contribution of inhomogeneity  to the entropy and heat capacity of
phases is calculated and the phase diagram in the coordinates
temperature\,--\,average concentration is constructed.
\newline%
\end{abstract}
\maketitle

\section{Introduction}\vspace{-0mm} 
The transition of binary solutions into a spatially inhomogeneous
state is observed in many systems, such as solid solutions
\cite{AGK}, $^3$He\,--\,$^4$He solutions \cite{BE} and solutions of
other liquids. A loss of stability (spinodal decomposition) has been
studied experimentally and theoretically for a long time in binary
solutions of normal liquids and polymers \cite{SS}. At present, more
and more attention is paid to the study of inhomogeneous media with
a complex internal structure, since it is in such systems that there
exists a real prospect of obtaining new materials with unique
properties. Therefore, the study of the problem of loss of stability
by many-particle physical systems and their transition into
spatially inhomogeneous states is of actual practical importance.
Such studies are also important from the point of view of
constructing a general theory of phase transitions into
inhomogeneous states. Although a large number of works have been
devoted to the phenomenon of spinodal decomposition, the theory of
not only dynamic phenomena but also transitions to equilibrium
states arising due to a loss of stability, all the more taking into
account nonlinear effects, still cannot be considered complete.

A phenomenological approach to the description of spinodal
decomposition was developed by Cahn and Hilliard \cite{CH1,CH2,CH3},
and also by Hillert \cite{MH}. A somewhat different formulation of
the theory was proposed in \cite{AGK}. The aim of this work is to
study the thermodynamics of the phase transition of a binary
solution into spatially inhomogeneous states with taking into
account nonlinear effects. The developed approach is basically close
to the works \cite{CH1,CH2,CH3,MH,AGK}, but differs in the    
implementation of the method. Since in practice the average
concentration in a system is initially determined by its
composition, in the proposed approach the average concentration is
considered to be a fixed parameter.

Based on the choice of a specific type of the free energy in the
form of an expansion in powers of the concentration fluctuations,
the character of phase transitions of binary solutions with
different average concentrations into spatially inhomogeneous states
is studied. The stratification of a solution and its subsequent
transition into the state with a concentration wave are
investigated. Thermodynamic potentials of inhomogeneous phases are
calculated and the type of transitions between phases is analyzed.
When calculating the entropy and heat capacity, the contribution of
all possible inhomogeneous states determined from the solution of
the nonlinear equation for concentration is taken into account.

\section{General thermodynamic relations }\vspace{-0mm} %
We will consider a solution of particles of two types, the number of
which is $N_1$ and $N_2$, occupying a constant volume $V$.  We
restrict ourselves to the isotropic approximation, assuming that in
a spatially homogeneous state the thermodynamics of such a solution
can be described by the free energy which is a function of
temperature $T,$ volume $V$ and the number of particles of each
component: $F=F(T,V,N_1,N_2)$. Sometimes it is more convenient to
pass to the total number of particles $N=N_1+N_2$ and the
concentration of one of the components $c_1\equiv\overline{c}=N_1/N$. 
As is known \cite{LL}, in a spatially homogeneous state the free
energy for the one-component case can be represented in the form $F=N\varphi(n,T)$. %
In the case of two components, a similar relation takes the form $F=N\varphi(n,T,c)$. %
In a spatially inhomogeneous state, we will consider the
concentration $c=c({\bf r})$ as a continuous function of spatial
coordinates. The total density $n$ is assumed to be constant.
However, it should be noted that in the case when the masses of
atoms or their radii differ greatly, then the observed effects can
be noticeably influenced by the change in density, which we do not
take into account for now. In this case, the total free energy will
be written in the form
\begin{equation} \label{01}
\begin{array}{l}
\displaystyle{%
   \overline{F}=n\!\int \!\varphi\big(T,n,\overline{c}\,; \delta c({\bf r}), \nabla\delta c({\bf r})\big)\, d{\bf r}.    %
}
\end{array}
\end{equation}
Thermodynamic potentials of this type are called potentials with
incomplete thermodynamic equilibrium. This means that temperature
and density are assumed to be equilibrium and independent of
coordinates, and the order parameters, in our case the deviation of
the concentration from the equilibrium mean value $\delta c({\bf
r})=c({\bf r})-\overline{c}$, should be found as a result of varying
the free energy (\ref{01}) with respect to the concentration from
the condition $\delta \overline{F}=0$, which gives the equation
\begin{equation} \label{02}
\begin{array}{l}
\displaystyle{%
   \frac{\partial \varphi}{\partial\delta c}-\nabla \frac{\partial \varphi}{\partial\nabla\delta c}=0. %
}
\end{array}
\end{equation}
Usually, the number of particles in a system is assumed to be
specified, since it is determined by the composition of a sample;
therefore, the average concentration $\overline{c}=N_1/N$ will be
considered to be a fixed parameter. By virtue of the requirement of
conservation of the total number of particles, there should hold the
conditions
\begin{equation} \label{03}
\begin{array}{l}
\displaystyle{%
  \frac{1}{V}\!\int \! \delta c({\bf r}; T,n,\overline{c})\, d{\bf r} =0,  \qquad   %
  \overline{c}=\frac{1}{V}\!\int \! c({\bf r}; T,n,\overline{c})\, d{\bf r}.   %
}%
\end{array}
\end{equation}
The solutions of Eq.\,(\ref{02}), depending on temperature, density
and the average concentration $\delta c=\delta c({\bf r}; T,n,\overline{c})$, %
should be substituted into the expression for the free energy
(\ref{01}). As a result, one obtains the formula for the equilibrium
free energy
\begin{equation} \label{04}
\begin{array}{l}
\displaystyle{%
   F=n\!\int \!\varphi\big(T,n,\overline{c}\,; \delta c({\bf r}; T,n,\overline{c}), \nabla\delta c({\bf r}; T,n,\overline{c})\big)\, d{\bf r}.    %
}
\end{array}
\end{equation}
In the following we choose the free energy density in the form
\begin{equation} \label{05}
\begin{array}{l}
\displaystyle{%
   \varphi=\frac{K}{2}\big(\nabla\delta c\big)^2 + \tilde{\varphi}(T,n,\overline{c}; \delta c)+\varphi_0(T,n,\overline{c}),  %
}
\end{array}
\end{equation}
where in the expansion in powers of the concentration fluctuations
we restrict ourselves, as is customary in the phenomenological
theory of phase transitions \cite{LL}, to terms no higher than the
fourth power including the cubic term, so that
\begin{equation} \label{06}
\begin{array}{l}
\displaystyle{%
   \tilde{\varphi}(T,n,\overline{c}; \delta c)=a_2\frac{(\delta c)^2}{2}+a_3\frac{(\delta c)^3}{3}+a_4\frac{(\delta c)^4}{4}.   %
}
\end{array}
\end{equation}
The last term $\varphi_0(T,n,\overline{c})$ in (\ref{05}) describes
the contribution of the homogeneous state to the total free energy.
The linear term in (\ref{06}) drops out due to the conservation
condition (\ref{03}). The coefficients in the expansion
(\ref{05}),\,(\ref{06}) depend, generally speaking, on temperature,
density, and also the average concentration
$a_i=a_i(T,n,\overline{c})$, $K=K(T,n,\overline{c})$. To ensure
stability of the homogeneous state in the region of high
temperatures, the coefficients $K$ and $a_4$ should be considered
positive, and the coefficient $a_3$ can be either positive or
negative. Let us assume that the coefficient $a_2$ changes sign at a
certain temperature, becoming negative at $T<T_p$. This spinodal
decomposition temperature is also a function of the average
concentration $T_p=T_p(\overline{c})$. In the following we will use
the linear approximation $a_2=a_0\tau,\,\,
\tau\equiv(T-T_p)\big/T_p\,\,(a_0>0)$. In the phenomenological
theory of phase transitions, such a dependence is used within the
framework of the mean field theory near the phase transition
temperature. We will not introduce this restriction in the proposed
model. Above $T_p$ at $\tau>0$, the spatially homogeneous state,
which we will call the {\it H}\,-phase, remains stable for an
arbitrary concentration. In the temperature range $T\leq T_p$ of
interest to us, the dimensionless relative temperature varies in the
range $-1<\tau\leq 0$. The use of such a simple approximation, as we
shall see, leads to reasonable results and is consistent with the
derivation from the microscopic theory. Note, however, that at low
temperatures close to $T=0$, or which is the same to $\tau=-1$, the
applicability of the classical description may be violated, since
here it is necessary to take into account quantum effects. In this
paper we restrict ourselves to only the classical consideration. For
the free energy density of the form (\ref{05}),\,(\ref{06}),
equation \!(\ref{02}) gives
\begin{equation} \label{07}
\begin{array}{l}
\displaystyle{%
   K\Delta\delta c=a_2\delta c+a_3(\delta c)^2+a_4(\delta c)^3.  %
}
\end{array}
\end{equation}
Among all possible solutions of this nonlinear equation, the
physically realizable ones are only those for which \linebreak $0\leq c({\bf r})\leq 1$. %

\section{Spatially inhomogeneous states}\vspace{-0mm} 
Let us consider the transitions of a binary solution into spatially
inhomogeneous states, assuming that the concentration can depend
only on one spatial coordinate $x$, which varies within the limits
$-L/2\leq x\leq L/2$. In this one-dimensional case the total free
energy (\ref{01}) is written in the form
\begin{equation} \label{08}
\begin{array}{l}
\displaystyle{%
   \overline{F}=nA\!\int \!dx\left[\frac{K}{2}\left(\frac{d\,\delta c}{dx}\right)^2+a_2\frac{(\delta c)^2}{2}+a_3\frac{(\delta c)^3}{3}+a_4\frac{(\delta c)^4}{4}\right],   %
}
\end{array}
\end{equation}
where $A$ is the area of a sample, the normal to which is parallel
to the axis $x$. Equation (\ref{07}) in this case takes the form
\begin{equation} \label{09}
\begin{array}{l}
\displaystyle{%
   K\frac{d^2\delta c}{dx^2}=a_2\delta c+a_3(\delta c)^2+a_4(\delta c)^3.  %
}
\end{array}
\end{equation}
The first integral of this equation is
\begin{equation} \label{10}
\begin{array}{l}
\displaystyle{%
   \frac{K}{2}\left(\frac{d\,\delta c}{dx}\right)^2=C-U(\delta c),   %
}
\end{array}
\end{equation}
where $C$ is the constant of integration, and
\begin{equation} \label{11}
\begin{array}{l}
\displaystyle{%
   U(\delta c)=-a_4\frac{(\delta c)^4}{4}-a_3\frac{(\delta c)^3}{3}-a_2\frac{(\delta c)^2}{2}.    %
}
\end{array}
\end{equation}
Equation (\ref{10}) is similar to the equation of motion of a
material point in an external field $U(\delta c)$ (\ref{11}) with
energy $C$. In further analysis it will sometimes be more convenient
to pass to dimensionless quantities, which we define by the
relations:
\begin{equation} \label{12}
\begin{array}{ccc}
\displaystyle{%
   c=\sqrt{g}\,\tilde{c},\qquad x\equiv\xi_0\tilde{x},       %
}\vspace{3mm}\\ %
\displaystyle{%
   g\equiv\frac{a_0}{a_4}>0,\quad b\equiv\frac{a_3}{g^{1\!/2}a_4}=g^{1\!/2}\frac{a_3}{a_0},   %
}\vspace{2mm}\\ %
\displaystyle{%
   \varepsilon\equiv\frac{C}{a_4g^2}=\frac{C}{a_0g}.  %
}%
\end{array}
\end{equation}
Here is introduced the correlation length $\xi_0$:
\begin{equation} \label{13}
\begin{array}{l}
\displaystyle{%
   \xi_0^2\equiv\frac{K}{ga_4}=\frac{K}{a_0}.   %
}
\end{array}
\end{equation}
In this notation Eq.\,(\ref{10}) and the ``field'' (\ref{11}) take the form %
\begin{equation} \label{14}
\begin{array}{l}
\displaystyle{%
   \frac{1}{2}\!\left(\frac{d\,\delta \tilde{c}}{d\tilde{x}}\right)^{\!2}+\tilde{U}(\delta \tilde{c})=\varepsilon,   %
}
\end{array}
\end{equation}
\vspace{-4mm}
\begin{equation} \label{15}
\begin{array}{l}
\displaystyle{%
   \tilde{U}(\delta \tilde{c})=-\frac{(\delta \tilde{c})^4}{4}-b\frac{(\delta \tilde{c})^3}{3}-\tau\frac{(\delta \tilde{c})^2}{2}.    %
}
\end{array}
\end{equation}
The use of the reduced concentration $\tilde{c}$, along with the
physical concentration $c$, is convenient because the dependence on
the positive parameter $g$ dropped out of the formula (\ref{15}). As
a result, the form of the field (\ref{15}) is determined by the
single dimensionless parameter $b=a_3\big/a_4g^{1\!/2}$ and also by
the dimensionless temperature $\tau$. For $b=0$ the field (\ref{15})
does not depend at all on material constants, but the parameter $g$
determines the region of permissible change in the reduced concentration %
\begin{equation} \label{16}
\begin{array}{l}
\displaystyle{%
   0\leq\tilde{c}\leq\frac{1}{\sqrt{g}}.   %
}
\end{array}
\end{equation}
The solution of Eq.\,(\ref{14}) in general form is given by the formula %
\begin{equation} \label{17}
\begin{array}{l}
\displaystyle{%
   \int\!\frac{d\,\delta\tilde{c}}{\sqrt{\varepsilon-\tilde{U}(\delta\tilde{c})}}=\pm\sqrt{2}\,\tilde{x}.   %
}
\end{array}
\end{equation}
An analysis of possible solutions of Eq.\,(\ref{17}) may be more
convenient in terms of the total concentration. In this
representation the field (\ref{15}) takes the form
\begin{equation} \label{18}
\begin{array}{l}
\displaystyle{%
   \tilde{U}(\tilde{c})=-\frac{1}{4}\left(\tilde{c}-\tilde{\overline{c}}\right)^2\left[\tilde{c}^{\,2} %
   +2\!\left(\!-\tilde{\overline{c}}\,+\frac{2}{3}b\right)\!\tilde{c}+\tilde{\overline{c}}^{\,2}\!-\!\frac{4}{3}b\,\tilde{\overline{c}}\,+2\tau \right].   %
}
\end{array}
\end{equation}

In order to demonstrate the application of the proposed method for
describing phase transitions into spatially inhomogeneous states, in
this work we restrict ourselves to considering the particular case
$b=0$. The conditions to which the crystal symmetry must conform for
this purpose in the case of solid solutions are analyzed, for
example, \linebreak in \cite{AGK}. The analysis of the more general
case $b\neq 0$ does not present any fundamental difficulty and can
be carried
out in a similar way.  For $b=0$, the field (\ref{18}) takes the form %
\begin{equation} \label{19}
\begin{array}{l}
\displaystyle{%
   \tilde{U}(\tilde{c})=-\frac{1}{4}\left(\tilde{c}-\tilde{\overline{c}}\right)^2 %
   \left[ \left(\tilde{c}-\tilde{\overline{c}}\right)^2 + 2\tau \right].   %
}
\end{array}
\end{equation}
If $\tau>0$, the field (\ref{19}) vanishes at
$\tilde{c}=\tilde{\overline{c}}$\, and takes negative values at all
other concentrations. Thus, in this temperature range the field
(\ref{19}) has no minimum and, therefore, only the homogeneous state
is possible. If $\tau<0$, then the function $\tilde{U}(\tilde{c})$
has the minimum $\tilde{U}(\tilde{\overline{c}})=0$ and the maximums
$\tilde{U}(\tilde{c}_{m1})=\tilde{U}(\tilde{c}_{m2})=|\tau|^2\!\big/4$ at %
\begin{equation} \label{20}
\begin{array}{l}
\displaystyle{%
   \tilde{c}_{m1}=\tilde{\overline{c}}\,-\!\sqrt{|\tau|},\qquad \tilde{c}_{m2}=\tilde{\overline{c}}\,+\!\sqrt{|\tau|}. %
}
\end{array}
\end{equation}
At concentrations
\begin{equation} \label{21}
\begin{array}{l}
\displaystyle{%
   \tilde{c}_{01}=\tilde{\overline{c}}\,-\!\sqrt{2|\tau|},\qquad \tilde{c}_{02}=\tilde{\overline{c}}\,+\!\sqrt{2|\tau|} %
}
\end{array}
\end{equation}
the field vanishes
$\tilde{U}(\tilde{c}_{01}\!)=\tilde{U}(\tilde{c}_{02}\!)=0$. Taking
into account the condition (\ref{16}) for concentration, there are
two qualitatively different possibilities, which are shown in Fig.\,1. %
At concentrations $0<\overline{c}\leq 1/2$ and temperatures
$\tau<-\,\overline{c}^{\,2}\!\big/g$ (Fig.\,1{\it a}), as well as
$1/2\leq\overline{c}< 1$ and $\tau<-(1-\overline{c})^{2}\!\big/g$ %
(Fig.\,1{\it b}), the maximum possible value of the parameter
$\varepsilon=\varepsilon_m$ is less than the value of the field at
the maximum. Here, as will be shown, there are only solutions in the
form of concentration waves. At\, $0<\overline{c}\leq 1/2$,\,\,$\tau\geq-\,\overline{c}^{\,2}\!\big/g$ %
and $1/2\leq\overline{c}< 1$,\,\,$\tau \geq -(1-\overline{c})^{2}\!\big/g$ (Fig.\,1{\it c}), %
the maximum possible value of the parameter
$\varepsilon=\varepsilon_m$ coincides with the value of the field at
the maximum. In this case the solution with the minimum energy at
$\varepsilon=\varepsilon_m$ has the form of a kink (31).
\vspace{-1mm} %
\begin{figure}[h!]
\vspace{-0mm}  \hspace{2mm}
\includegraphics[width = 0.94\columnwidth]{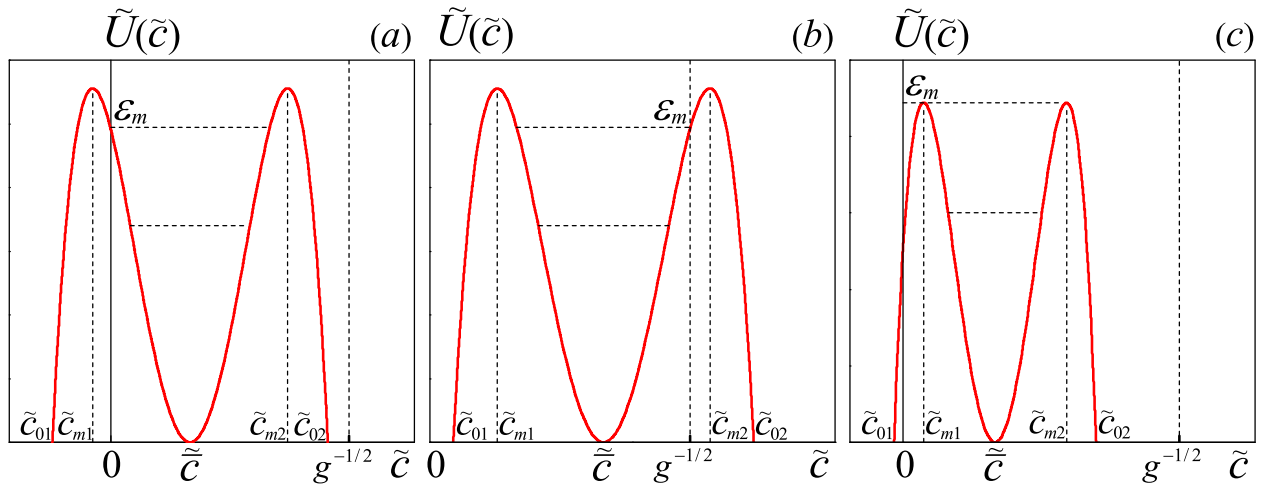} 
\vspace{-3mm} %
\caption{\label{fig01} 
The form of the field $\tilde{U}(\tilde{c})$ at different values of average concentration and temperature.  %
(\!{\it a})   $0<\overline{c}\leq 1/2$,\,\,$\tau<-\,\overline{c}^{\,2}\!\big/g$,\, %
(\!{\it b})   $1/2\leq\overline{c}< 1$,\,\,$\tau<-(1-\overline{c})^{2}\!\big/g$\,\,\,(\!{\it W}\,-phase),\, %
(\!{\it c})   $0<\overline{c}\leq 1/2$,\,\,$\tau\geq-\,\overline{c}^{\,2}\!\big/g$; %
$1/2\leq\overline{c}< 1$,\,\,$\tau\geq-(1-\overline{c})^{2}\!\big/g$\,\,\,({\it K}\,-phase). %
}%
\end{figure}

Thus, along with the homogeneous {\it H}\,-phase, the stable
inhomogeneous {\it K}\,-\,  and {\it W}\,-\,phases can exist in binary solutions. %

\section{Free energy parameters}\vspace{0mm} %
Further consideration will be based on the expansion of the free
energy in powers of the concentration fluctuations (\ref{08}). To
estimate the phenomenological parameters entering in
(\ref{05}),\,(\ref{06}) and find their dependence on the average
concentration, temperature and density, we use the well-known
representation of the free energy \cite{AGK,MT}, which in the
considered case of an isotropic medium can be represented as
\begin{equation} \label{22}
\begin{array}{l}
\displaystyle{%
   \overline{F}=\frac{n^2}{2}\!\int \! d{\bf r}d{\bf r}' V\big(|{\bf r}-{\bf r}'|\big)c({\bf r})c({\bf r}') + %
   nT\!\int \! d{\bf r}\Big[c({\bf r})\ln c({\bf r}) + \big(1-c({\bf r})\big)\ln \big(1-c({\bf r})\big) \Big],  %
}
\end{array}
\end{equation}
where $V\big(|{\bf r}-{\bf r}'|\big)$ is the interaction potential
between particles. Expanding (\ref{22}) in small fluctuations %
$\delta c({\bf r})=c({\bf r})-\overline{c}$ and comparing with
(\ref{04})\,--\,(\ref{06}), we obtain the following expressions for
the expansion coefficients:
\begin{equation} \label{23}
\begin{array}{ccc}
\displaystyle{%
   a_2=nV_0+\frac{T}{\overline{c}\,(1-\overline{c})},\qquad  a_3=-\frac{(1-2\overline{c})}{2\,\overline{c}^{\,2}(1-\overline{c})^2}\,T,\qquad    %
   a_4=\frac{(1-3\overline{c}+3\overline{c}^{\,2})}{3\,\overline{c}^{\,3}(1-\overline{c})^3}\,T, %
}\vspace{2mm}\\ %
\displaystyle{%
   K=-\frac{2\pi}{3}\,n\!\int \! dr\,r^4V(r), \qquad %
   \varphi_0(T,\overline{c})=\frac{\overline{c}^{\,2}n}{2}\,V_0 + T\big[\,\overline{c}\ln \overline{c} + (1-\overline{c})\ln (1-\overline{c})\big], %
}%
\end{array}
\end{equation}
where $V_0=\!\int d{\bf r}V(r)$. When calculating the gradient
contribution, we used the expansion %
$\delta c({\bf r}+{\boldsymbol{\uprho}})\approx\delta c({\bf r})+\rho_i\nabla_i\,\delta c({\bf r})+\frac{1}{2}\rho_i\rho_k\nabla_i\nabla_k\,\delta c({\bf r})$ %
in the first term in (\ref{22}), and also took into account that in
an isotropic medium
\begin{equation} \label{24}
\begin{array}{ccc}
\displaystyle{%
   \int\!d{\boldsymbol{\uprho}}\,\rho_i V(\rho) = 0, \qquad %
   \int\!d{\boldsymbol{\uprho}}\,\rho_i\rho_k V(\rho) = \delta_{ik}\,\frac{1}{3}\int\! d\rho\,\rho^2 V(\rho). %
}%
\end{array}
\end{equation}
As is seen, the coefficient $a_2$ can change sign at a certain
temperature, becoming negative if only $V_0<0$, so that the
possibility of a loss of stability is determined by the character of
the interaction between atoms. In the potential of interparticle
interaction, as a rule, it is possible to distinguish the region of
repulsion at small distances and the region of attraction at large
distances. For many model potentials it is assumed that the
repulsion at small distances becomes infinitely large, which turns
out to be inconvenient in theoretical studies. However, it is quite
admissible and natural to consider effective potentials that remain
finite at small distances \cite{YP1,PS}. For further estimates we
will use the modified Sutherland potential
\begin{equation} \label{25}
\begin{array}{ll}
\displaystyle{%
  V(r)=\left\{
               \begin{array}{l}
                 \,\,\,\,v_0,\qquad\qquad\quad                   r<r_0, \vspace{3mm}  \\
                \displaystyle{
                 -\,v_m\!\left(\frac{r_0}{r}\right)^{\!6}, \quad\,\,\,\, r>r_0, %
                 }
               \end{array} \right.
}%
\end{array}
\end{equation}
where $v_0>0$ and $v_m>0$. For this potential we obtain
\begin{equation} \label{26}
\begin{array}{l}
\displaystyle{%
   V_0=\frac{4\pi}{3}r_0^3(v_0-v_m), \qquad    K=-\frac{2\pi}{15}\,nr_0^5(v_0-5v_m). %
}%
\end{array}
\end{equation}
For a sufficiently deep well in the potential (\ref{25}) $v_m>v_0$,
there holds the condition $V_0<0$. In this case it is also ensured
the condition of stability of the homogeneous state $K>0$. The
spinodal decomposition temperature is then determined by the formula
\begin{equation} \label{27}
\begin{array}{l}
\displaystyle{%
   T_p=n|V_0|\,\overline{c}\,(1-\overline{c}).    %
}%
\end{array}
\end{equation}
In the coefficient $a_2=a_0\tau$ the proportionality parameter is
\begin{equation} \label{28}
\begin{array}{l}
\displaystyle{%
   a_0\equiv\frac{T_p}{\overline{c}\,(1-\overline{c})}=n|V_0|.    %
}%
\end{array}
\end{equation}
Let us also present expressions for the dimensionless parameters (\ref{12}): %
\begin{equation} \label{29}
\begin{array}{ccc}
\displaystyle{%
   g\equiv g(\tau,\overline{c})=\frac{g_0(\overline{c})}{(1+\tau)}, \qquad g_0(\overline{c})\equiv\frac{3\overline{c}^{\,2}(1-\overline{c})^2}{(1-3\overline{c}+3\overline{c}^{\,2})},     %
}\vspace{2mm}\\ %
\displaystyle{%
   b\equiv -(1+\tau)^{1\!/2}b_0(\overline{c}), \qquad  b_0(\overline{c})\equiv\frac{\sqrt{3}}{2}\frac{(1-2\overline{c})}{(1-3\overline{c}+3\overline{c}^{\,2})^{1\!/2}}.    %
}%
\end{array}
\end{equation}
For completeness we gave a formula for the coefficient $b$,
although, as was said, it is omitted in this work. The correlation
length for potential (\ref{25}) is given by the formula
\begin{equation} \label{30}
\begin{array}{ccc}
\displaystyle{%
   \xi_0^2=\frac{2\pi}{3V_0}\!\int \! d\rho\,\rho^4V(\rho)=r_0^2\frac{(v_m-v_0/5)}{2(v_m-v_0)},  %
}%
\end{array}
\end{equation}
and is of the order of magnitude of the radius of action of the
interparticle potential $\xi_0\sim r_0$. The macroscopic description
and expansion in terms of small concentration gradients is valid if
the characteristic distance at which the concentration changes $L_c$
is much larger than the correlation length $L_c\gg\xi_0$. We also
note that to analyze nonlinear effects we use the expansion in
powers of concentration (\ref{08}), without imposing restrictions on
the magnitude of the concentration fluctuations.

\section{Phase diagram}\vspace{-0mm} %
Bounded solutions of Eq.\,(\ref{17}) exist in the range of variation
of the parameter $0\leq\varepsilon\leq\varepsilon_m$. As noted
above, depending on the values of $\tau$ and $\overline{c}$ there
are two possibilities. First, at permissible concentration values
the parameter $\varepsilon$ can take its maximum value equal to the
field maximum $\varepsilon_m=\tilde{U}(\tilde{c}_{m1})=\tilde{U}(\tilde{c}_{m2})=|\tau|^2\!\big/4$ %
(Fig.\,1{\it c}). In this case, at $\varepsilon=\varepsilon_m$ the
solution of Eq.\,(\ref{17}) has the form of a kink
\begin{equation} \label{31}
\begin{array}{ccc}
\displaystyle{%
   \delta\tilde{c}(\tilde{x})=\tilde{c}(\tilde{x})-\tilde{\overline{c}}=\pm\sqrt{|\tau|}\,\,{\rm th}\!\!\left(\!\frac{\sqrt{|\tau|}\,\tilde{x}}{\sqrt{2}}\right).  %
}%
\end{array}
\end{equation}
This formula describes a solution, in which with a change of the
coordinate a transition occurs from the homogeneous state with an
increased concentration to the homogeneous state where the
concentration is less than the average one, so that %
$c(+\infty)-c(-\infty)=2\sqrt{g|\tau|}$. The solution stratifies
into two areas, in which the concentration is above and below the
average one. The width of the transition region, where the
concentration changes significantly, is given by the formula %
$L_K=\sqrt{\frac{2}{|\tau|}}\,\xi_0$. At $\tau\rightarrow 0$ the
concentration difference tends to zero and the width of the
transition region tends to infinity, so that in the result the
system goes over to the spatially homogeneous state.

In the range of variation of the parameter
$0<\varepsilon<\varepsilon_m$, the solution has the form of a
concentration wave
\begin{equation} \label{32}
\begin{array}{ccc}
\displaystyle{%
   \delta\tilde{c}(\tilde{x})=\tilde{c}(\tilde{x})-\tilde{\overline{c}}=\pm\, D_- \,{\rm sn}\!\left(\!\frac{\tilde{x}}{\sqrt{2}}\,D_+,m \right),  %
}%
\end{array}
\end{equation}
where $\displaystyle{D_\pm\equiv D_\pm(\tau,\varepsilon)\equiv\sqrt{|\tau|\pm\sqrt{\tau^2-4\varepsilon}}}$, %
$m\equiv m(\tau,\varepsilon)=D_-^2\big/D_+^2$, ${\rm sn}\!\left(\!\frac{\tilde{x}D_+}{\sqrt{2}},m \right)$ %
is the elliptic sine function \cite{AS}. The description of the
distribution of atoms of a solid solution by means of concentration
waves was proposed by Krivoglaz \cite{MK} and developed in relation
to the description of ordered solutions in \cite{AGK}. In the limit
$\varepsilon\rightarrow|\tau|^2\!\big/4$ or $m\rightarrow 1$, this
solution goes over to the kink solution (\ref{31}). The case
$\varepsilon=0$ corresponds to the spatially homogeneous state. The
solution with the maximum possible value $\varepsilon=\varepsilon_m$
in the form of a kink (\ref{31}) corresponds to the minimum value of
the free energy.

In another possible case, the maximum value of the parameter $\varepsilon=\varepsilon_m$ %
at all permissible concentrations is less than the maximum value of
the field $\varepsilon_m<|\tau|^2\!\big/4$ (Fig.\,1{\it a},{\it b}).
At $\overline{c}\leq 1/2$\hspace{3mm}$\displaystyle{\varepsilon_m=\tilde{U}(0,\tau)=-\frac{\overline{c}^{\,2}}{4g}\left(\frac{\overline{c}^{\,2}}{g}+2\tau\right)}$, and %
at $\overline{c}\geq 1/2$\hspace{3mm}$\displaystyle{\varepsilon_m=\tilde{U}\left(\frac{1}{\sqrt{g}},\tau\right)=-\frac{(1-\overline{c})^2}{4g}\left(\frac{(1-\overline{c})^2}{g}+2\tau\right)}$. %
In this case, there is no solution in the form of a kink and all
bounded solutions at $0<\varepsilon\leq\varepsilon_m$ have the form
of concentration waves (\ref{32}).

\vspace{0mm} %
\begin{figure}[b!]
\vspace{-0mm}  \hspace{-10mm}
\includegraphics[width = 8.1cm]{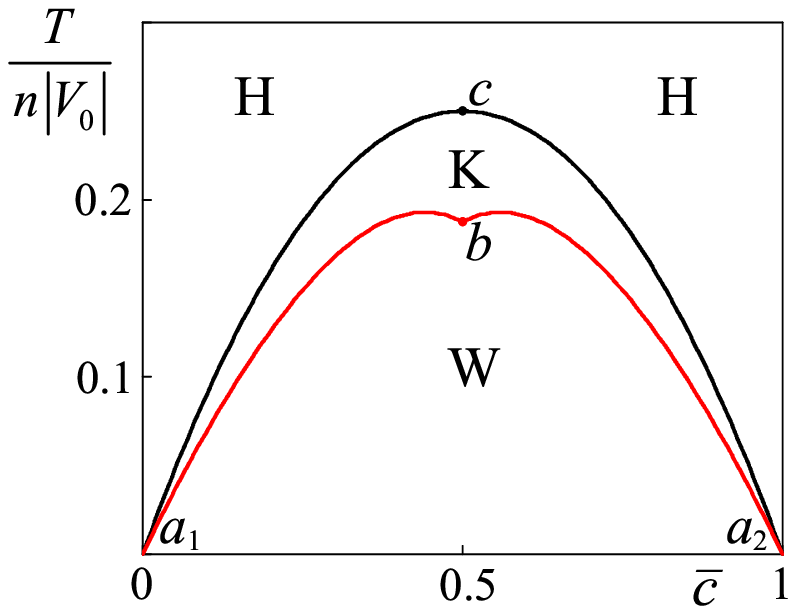} 
\vspace{-3mm} %
\caption{\label{fig02} 
Phase diagram in the coordinates $\big(\overline{c},T\!\big/n|V_0|\big)$. %
The lines $a_1ca_2$, $a_1b$, $ba_2$ respectively correspond to the dependencies: %
$T_p\big/n|V_0|=\overline{c}\,(1-\overline{c})$, %
$T_1\!\big/n|V_0|=3\overline{c}\,(1-\overline{c})^{3}\!\big/(4-9\overline{c}+6\overline{c}^{\,2})$, %
$T_2\big/n|V_0|=3\overline{c}^{\,3}(1-\overline{c})\!\big/(1-3\overline{c}+6\overline{c}^{\,2})$. %
Above the curve $a_1ca_2$ there is the homogeneous {\it H}\,-phase.
Inside the area $a_1ca_2b$ is the {\it K}\,-phase. Inside the area
$a_1ba_2$ is the {\it W}\,-phase.
}%
\end{figure}

As we can see, while at $\tau\geq 0$ there is only the spatially
homogeneous {\it H}\,-phase, at $\tau<0$ there are two qualitatively
different states. In one of them, in addition to the solutions in
the form of concentration waves, there exists the solution in the
form of a kink (\ref{31}), which actually corresponds to the
smallest value of the free energy. We called this state the {\it
K}\,-phase. In this phase the solution is stratified into two
spatial regions with increased and depleted concentrations. In the
other phase all solutions have the form of concentration waves
(\ref{32}), and there is no solution in the form of a kink. We
called this state the {\it W}\,-phase. The phase diagram in the
coordinates temperature\,--\,average concentration is shown in
Fig.\,2. Note that at $\overline{c}\rightarrow 0$ we have
$T_p\sim\overline{c}$, and at $\overline{c}\rightarrow 1$ we have
$T_p\sim(1-\overline{c})$, so that the transition temperature tends
to zero in both limiting cases. Here, generally speaking, due to the
proximity of $T_p$ to absolute zero the applicability of the model
and the used classical description may be violated.

The concentration distributions in the inhomogeneous phases during
stratification ({\it a}) and in the concentration wave ({\it b}) are
shown in Fig.\,3. As is seen, transitions between the phases are
accompanied by the transfer of a significant amount of matter. In
this work, however, we do not consider the kinetics of transitions
between the phases and do not estimate the characteristic times
during which such transitions occur, but we study only
thermodynamically equilibrium states.

\vspace{0mm} %
\begin{figure}[t!]
\vspace{-0mm}  
\includegraphics[width = 15.5cm]{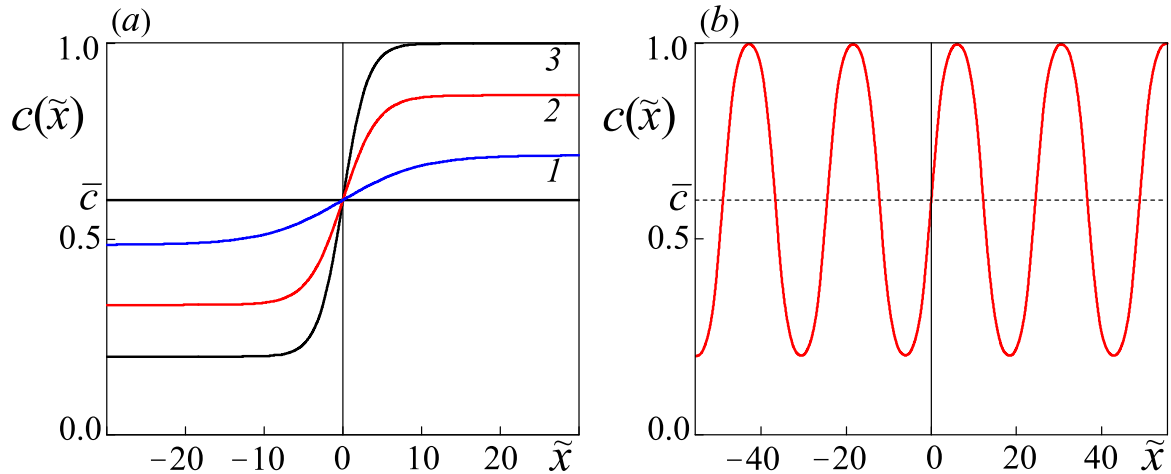} 
\vspace{-3mm} %
\caption{\label{fig03} 
(\!{\it a}) The concentration distributions $c(\tilde{x})_K$ in the
{\it K}\,-phase ($\tilde{T}_2\leq\tilde{T}<\tilde{T}_p=0.24$) %
at temperatures $\tilde{T}=T\!\big/n|V_0|$\,: %
{\it 1}\,--\,\,$\tilde{T}=0.235$, {\it 2}\,--\,\,$\tilde{T}=0.215$, %
{\it 3}\,--\,\,$\tilde{T}_2=0.191$. %
The line $c=\overline{c}=0.6$ corresponds to the {\it H}\,-phase at
$\tilde{T}\geq\tilde{T}_p$. (\!{\it b}) The concentration
distribution $c(\tilde{x})_W$ in the {\it W}\,-phase $(\tilde{T}<\tilde{T}_2)$ %
at $\tilde{T}=0.185$.
}%
\end{figure}

\section{Thermodynamics of inhomogeneous states}\vspace{-0mm} %
Let us calculate the contribution of inhomogeneity to the
thermodynamic functions of the phases and analyze the character of
transitions between the phases. Substituting the solution at a
certain value of $\varepsilon$ into (\ref{08}), we find the
contribution of this state to the equilibrium free energy. For the
kink in the limit $L\rightarrow\infty$, we obtain
\begin{equation} \label{33}
\begin{array}{ccc}
\displaystyle{%
   \frac{F}{a_4g^2N}=-\varepsilon=-\frac{\tau^2}{4}.  %
}%
\end{array}
\end{equation}
For the concentration wave at a certain value of $\varepsilon$, we have %
\begin{equation} \label{34}
\begin{array}{ccc}
\displaystyle{%
   \frac{F(\tau,\varepsilon)}{a_4g^2N}=\frac{1}{3}\!\left[\varepsilon+\tau D_+^2(\tau,\varepsilon)\left(1-\frac{E(m)}{K(m)}\right)\right],  %
}%
\end{array}
\end{equation}
where $K(m), E(m)$ are the complete elliptic integrals of the first
and second kind \cite{AS}. In the following, in (\ref{34}) it is more %
convenient to express the parameter $\varepsilon$ in terms of the parameter %
\begin{equation} \label{35}
\begin{array}{ccc}
\displaystyle{%
    m=\frac{|\tau|-\sqrt{\tau^2-4\varepsilon}}{|\tau|+\sqrt{\tau^2-4\varepsilon}},  %
}%
\end{array}
\end{equation}
which varies from $m=0$ at $\varepsilon=0$ to %
$m=m_*(\tau)\equiv\displaystyle{\Big(|\tau|-\sqrt{\tau^2-4\varepsilon_m}\Big)\!\Big/\!\Big(|\tau|+\sqrt{\tau^2-4\varepsilon_m}}\Big)$
at $\varepsilon=\varepsilon_m$. So we find the energy of the
inhomogeneous state per one particle
\begin{equation} \label{36}
\begin{array}{ccc}
\displaystyle{%
   \frac{F(\tau,m)}{N}=-G\tau^2f(m), %
}%
\end{array}
\end{equation}
where the function
\begin{equation} \label{37}
\begin{array}{ccc}
\displaystyle{%
   f(m)\equiv\frac{(2+m)}{(1+m)^2}-\frac{2}{(1+m)}\frac{E(m)}{K(m)} %
}%
\end{array}
\end{equation}
is defined, and $G\equiv\frac{1}{3}a_4g^2$  is a constant which with
taking into account the formulas
(\ref{23}),\,(\ref{27}),\,(\ref{29}) can be presented in the form
\begin{equation} \label{38}
\begin{array}{ccc}
\displaystyle{%
   G=T_p\frac{\eta(\overline{c})}{(1+\tau)},\qquad  %
   \eta(\overline{c})\equiv\frac{\overline{c}(1-\overline{c})}{(1-3\overline{c}+3\overline{c}^{\,2})}.    %
}%
\end{array}
\end{equation}
The form of the function (\ref{37}) is shown in Fig.\,4. It is
positive and increases from $f(0)=0$ to $f(1)=3/4$, and a very rapid
growth occurs at $m\approx 1$, so that here the derivative goes to infinity. %
\vspace{0mm} %
\begin{figure}[h!]
\vspace{-0mm}  \hspace{-20mm}
\includegraphics[width = 8.0cm]{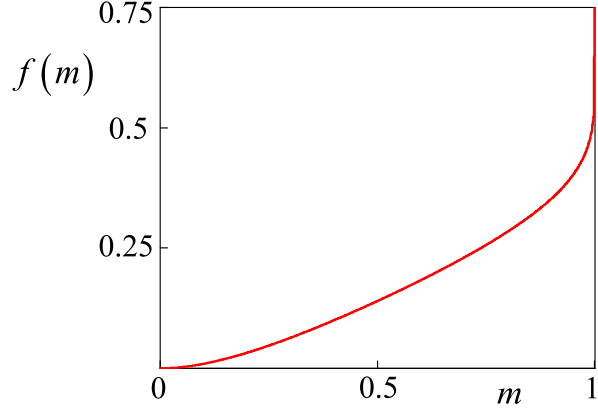} 
\vspace{-3mm} %
\caption{\label{fig04} 
Form of the function $f(m)$ (\ref{37}).
}%
\end{figure}

The function $F(\tau,m)$ (\ref{36}) determines the contribution of
solution with index $m$ to the energy. If excitations with different
energies are possible in a physical system, then all of them
contribute to the total equilibrium free energy. According to the
general principles of statistical physics, the probability of
finding the system in the state with the parameter $m$ in the
interval $m\div m+dm$
\begin{equation} \label{39}
\begin{array}{ccc}
\displaystyle{%
   dw=\phi(T,m)\,dm       %
}%
\end{array}
\end{equation}
is determined by means of the ``one-particle'' distribution function
over states $m$
\begin{equation} \label{40}
\begin{array}{ccc}
\displaystyle{%
   \phi(T,m)\equiv\frac{e^{ -\frac{F(\tau,m)}{NT} }}{Z}=\frac{e^{B(\tau)f(m)}}{Z}, %
}%
\end{array}
\end{equation}
where we use the notation
\begin{equation} \label{41}
\begin{array}{ccc}
\displaystyle{%
   B(\tau)\equiv\frac{\tau^2}{(1+\tau)^2}\,\eta(\overline{c}).   %
}%
\end{array}
\end{equation}
The statistical integral is determined by the normalization condition %
$\int_0^{m_*(\tau)}\!\!\phi(T,m)\,dm=1$:
\begin{equation} \label{42}
\begin{array}{ccc}
\displaystyle{%
   Z=\int\limits_0^{m_*(\tau)}e^{B(\tau)f(m)}dm.   %
}%
\end{array}
\end{equation}
The average of an arbitrary function $A=A(m)$ is given by the formula %
\begin{equation} \label{43}
\begin{array}{ccc}
\displaystyle{%
   \big\langle A\big\rangle=\int\limits_0^{m_*(\tau)}A(m)\phi(\tau,m)dm.   %
}%
\end{array}
\end{equation}
As follows from the form of the function (\ref{37}) (Fig.\,4), the
largest contribution is made by the state with the maximum value
$m=m_*$, while the contribution of the states with smaller values of
$m$ rapidly decreases.

The average value of the function (\ref{36}) determines the
contribution of inhomogeneous states to the average energy
\begin{equation} \label{44}
\begin{array}{ccc}
\displaystyle{%
  E=\int\limits_0^{m_*}F(\tau,m)\phi(\tau,m)dm=-NG\,\tau^2\big\langle f\big\rangle.   %
}%
\end{array}
\end{equation}
The contribution to the total entropy due to inhomogeneity of states
is determined through the average of the logarithm of the
distribution function \cite{LL}, so that the entropy per one
particle is given by the formula
\begin{equation} \label{45}
\begin{array}{ccc}
\displaystyle{%
  \frac{S}{N}\equiv -\big\langle \ln\phi\,\big\rangle=\ln Z-B\big\langle f\big\rangle.   %
}%
\end{array}
\end{equation}
The contribution of inhomogeneity to the total heat capacity can be
determined through the derivative of the entropy
\begin{equation} \label{46}
\begin{array}{ccc}
\displaystyle{%
  \frac{C}{N}=\frac{T}{N}\frac{dS}{dT}=\frac{2\,\tau^3}{(1+\tau)^4}\,\eta^2(\overline{c})\Big(\!\big\langle f\big\rangle^2-\big\langle f^{\,2}\big\rangle\!\Big)+   %
  (1+\tau)\frac{e^{Bf(m_*)}}{Z}\!\left[1+\frac{\tau^2}{(1+\tau)^2}\,\eta(\overline{c})\Big(\!\big\langle f\big\rangle-f(m_*)\Big)\right]\!\frac{dm_*}{d\tau}\,, %
}%
\end{array}
\end{equation}
where
\begin{equation} \label{47}
\begin{array}{ccc}
\displaystyle{%
   \big\langle f^{\,n} \big\rangle=\int\limits_0^{m_*}\,\big[f(m)\big]^n\phi(\tau,m)\,dm.   %
}%
\end{array}
\end{equation}
In particular, in the {\it K}\,-phase\, $m_*=1$, and in the {\it W}\,-phase %
\begin{equation} \label{48}
\begin{array}{ccc}
\displaystyle{%
   m_*(\tau)=\frac{(\tau+1)}{\tau\!\left(\!\displaystyle{\frac{2}{\tau_i}}+1\!\right)\!-1},\qquad  %
   \frac{dm_*}{d\tau}=-\frac{2\,(\tau_i+1)}{\tau_i\!\left[\tau\!\left(\!\displaystyle{\frac{2}{\tau_i}}+1\!\right)\!-1\right]^{\!2}}.    %
}%
\end{array}
\end{equation}
Here at $0<\overline{c}\leq 1/2$\,\, $i=1$, and at $1/2\leq\overline{c}<1$\,\, $i=2$, and at that  %
\begin{equation} \label{49}
\begin{array}{ccc}
\displaystyle{%
   \tau_1=\frac{T_1-T_p}{T_p}=-\frac{\overline{c}^{\,2}}{\left[\overline{c}^{\,2}+g_0(\overline{c})\right]}, \qquad %
   \tau_2=\frac{T_2-T_p}{T_p}=-\frac{ (1-\overline{c})^{2} }{\left[(1-\overline{c})^{2}+g_0(\overline{c})\right]}. %
}%
\end{array}
\end{equation}
The temperature dependences of the contributions of inhomogeneous
states to the entropy and heat capacity are shown in Fig.\,5.
\vspace{0mm} %
\begin{figure}[h!]
\vspace{-0mm}  
\includegraphics[width = 15.6cm]{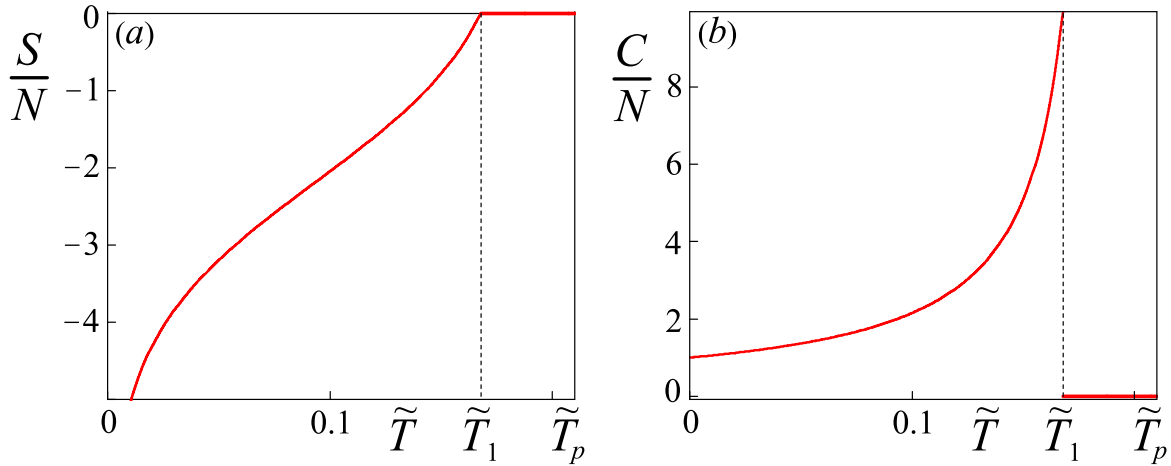} 
\vspace{-3mm} %
\caption{\label{fig05} 
Dependencies of (\!{\it a}) the entropy and (\!{\it b}) the heat
capacity on the dimensionless temperature $\tilde{T}=T\!\big/n|V_0|$
\newline ($\overline{c}=0.3$, $\tilde{T}_1=0.168$, $\tilde{T}_p=0.21$).
}%
\end{figure}

\noindent As we can see, in the {\it K}\,-phase at $\tilde{T}_1\leq\tilde{T}<\tilde{T}_p$ %
the entropy and heat capacity associated with inhomogeneity are
small, so that on the scale of the figure they are practically close
to zero. Upon transition to the {\it W}\,-phase the entropy
decreases, and the heat capacity at the transition temperature
$\tilde{T}_1$ increases by a jump and also decreases monotonically
with decreasing temperature. As it was noted, near absolute zero the
applicability of the used model and the classical approach may be violated. %
\newpage

The existence, along with the main solutions determining the
structure of the phases, also of other solutions in the form of
concentration waves with energies greater than the energy of the
main solution, will somewhat affect the concentration distribution
in the phases, but due to the randomness of phases this will not
lead to a qualitative change in the form of the concentration
distributions in the {\it K}\,-\, and {\it W}\,-phases shown in Fig.\,3. %

It is important to note that a loss of stability of the homogeneous
state is described by the solution of the nonlinear equation in the
form of a kink (\ref{31}) even near $T_p$. Thus, to analyze a loss
of stability it is not enough applying a linear approximation, and
taking into account nonlinear effects proves to be important even
for small $\tau$.

\section{Phase transition from the {\it H}\,-phase to the {\it K}\,-phase}\vspace{-2.5mm} %
As can be seen from the phase diagram in Fig.\,2, with decreasing
temperature at $T=T_p$ the solution loses its stability and begins
to stratify passing to the {\it K}\,-phase. Let us consider in more
detail the thermodynamics of this transition. Near the transition
from the homogeneous {\it H}\,-phase to the {\it K}\,-phase, where
$|\tau|\ll 1$ and $B\approx\eta(\overline{c})\tau^2<1$, from
(\ref{45}) and (\ref{46}) there follow formulas for the entropy and heat capacity: %
\begin{equation} \label{50}
\begin{array}{ccc}
\displaystyle{%
  \frac{S}{N}=\frac{\big[\eta(\overline{c})\big]^2}{2}\big(f_1^{\,2}-f_2\big)\hspace{0.3mm}\tau^4,  %
}%
\end{array}
\end{equation}\vspace{-3mm} %
\begin{equation} \label{51}
\begin{array}{ccc}
\displaystyle{%
  \frac{C}{N}=2\big[\eta(\overline{c})\big]^2\big(f_1^{\,2}-f_2\big)\hspace{0.3mm}\tau^3.  %
}%
\end{array}
\end{equation}
Here $f_n=\int_0^{1}\!\big[f(m)\big]^n dm$ and %
$f_1\approx 0.162$, $f_2\approx 0.043$, $f_1^{\,2}-f_2=-0.017$. %
Thus, there occurs a smooth phase transition from the homogeneous
state to the phase where the solution is stratified. In this case in
the inhomogeneous state the entropy slightly decreases and the heat
capacity increases (Fig.\,5). Only the fourth derivative of the
entropy with respect to temperature experiences a jump. The form of
the temperature dependences (\ref{50}) and (\ref{51}) is mainly
determined by the fact that the linear approximation of the
dependence of the coefficient $a_2=a_0(T-T_p)/T_p$ %
in (\ref{06}) on temperature was used. As is known \cite{LL,HS,PP},
near the phase transition temperature, fluctuations begin to play an
important role, which are not taken into account in this approach
based on the mean field theory. An effective account for
fluctuations should lead to a modification of the critical exponents
and temperature dependences of quantities near the phase transition \cite{GS,YP2}. %
\vspace{0mm} %
\begin{figure}[b!]
\vspace{-0mm}  
\includegraphics[width = 9.0cm]{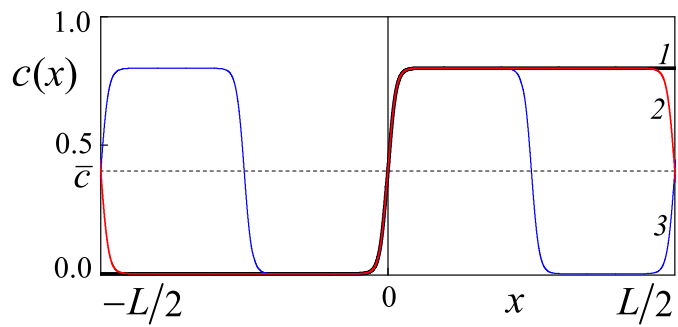} 
\vspace{-3mm} %
\caption{\label{fig06} 
Evolution of the concentration distribution during the transition of
the stratified solution into the phase with concentration waves.
Curve {\it 1} is the distribution at the transition boundary at $\tilde{T}=\tilde{T}_1$. %
Curves {\it 2} and {\it 3} are the distributions in the {\it W}\,-phase at %
$\tilde{T}<\tilde{T}_1$, with respectively one and two wave periods
at the sample length $L$. Parameters: $\overline{c}=0.4,\, \tilde{T}_1=0.191$. %
}%
\end{figure}

\section{Phase transition between inhomogeneous phases}\vspace{-2.5mm} %
As it was noted earlier (Fig.\,2), below the temperature $T_p$ after
a loss of stability of the homogeneous phase there first appears the
{\it K}\,-phase and the stratification of the solution occurs. The
phase with stratification exists in the temperature range %
$T_1\leq T<T_p$ if $\overline{c}\leq 1/2$, and
$T_2\leq T<T_p$ if $\overline{c}\geq 1/2$, where %
\begin{equation} \label{52}
\begin{array}{ccc}
\displaystyle{%
   T_1=T_p\frac{3(1-\overline{c})^2}{(4-9\overline{c}+6\overline{c}^{\,2})},\qquad  %
   T_2=T_p\frac{3\overline{c}^{\,2}}{(1-3\overline{c}+6\overline{c}^{\,2})}.  %
}%
\end{array}
\end{equation}
At lower temperatures $0<T<T_1$ if\, $\overline{c}\leq 1/2$, and
$0<T<T_2$ if\, $\overline{c}\geq 1/2$, the {\it W}\,-phase appears
(Fig.\,2). Figure 6 shows the evolution of the concentration
distribution near the transition from the stratified state to the
state with a concentration wave. With a further decrease in
temperature the concentration wave takes a shape close to \linebreak
sinusoidal (Fig.\,3{\it b}).\newpage

Let us consider in more detail the character of the transition from
the {\it K}\,-phase to the {\it W}\,-phase. In the region of
existence of the {\it K}\,-phase the entropy and heat capacity can
be represented in the form
\begin{equation} \label{53}
\begin{array}{ccc}
\displaystyle{%
  \frac{S_K(\tau)}{N}=\ln Z_K-\frac{\tau^2}{(1+\tau)^2}\,\eta(\overline{c})\big\langle f\big\rangle_{\!K},   %
}%
\end{array}
\end{equation}
\vspace{-3mm}
\begin{equation} \label{54}
\begin{array}{ccc}
\displaystyle{%
  \frac{C_K(\tau)}{N}=\frac{2\,\tau^3}{(1+\tau)^4}\,\eta^2(\overline{c})\Big(\!\big\langle f\big\rangle_{\!K}^2-\big\langle f^{\,2}\big\rangle_{\!K}\!\Big),   %
}%
\end{array}
\end{equation}
where
\begin{equation} \label{55}
\begin{array}{ccc}
\displaystyle{%
   Z_K=\int\limits_0^{1}\!e^{B f(m)}dm, \qquad %
   \big\langle f\big\rangle_{\!K}=\frac{1}{Z_K}\int\limits_0^{1}\!f(m)\hspace{0.5mm}e^{B f(m)}dm, \qquad  %
   \big\langle f^{\,2}\big\rangle_{\!K}=\frac{1}{Z_K}\int\limits_0^{1}\!f^{\,2}(m)\hspace{0.5mm}e^{B f(m)}dm.  %
}%
\end{array}
\end{equation}
In the {\it W}\,-phase with concentration waves the same quantities have the form %
\begin{equation} \label{56}
\begin{array}{ccc}
\displaystyle{%
  \frac{S_W(\tau)}{N}=\ln Z_W-\frac{\tau^2}{(1+\tau)^2}\,\eta(\overline{c})\big\langle f\big\rangle_{\!W},   %
}%
\end{array}
\end{equation}
\vspace{-3mm}
\begin{equation} \label{57}
\begin{array}{ccc}
\displaystyle{%
  \frac{C_W(\tau)}{N}=\frac{2\,\tau^3}{(1+\tau)^4}\,\eta^2(\overline{c})\Big(\!\big\langle f\big\rangle_{\!W}^2-\big\langle f^{\,2}\big\rangle_{\!W}\!\Big)+   %
  (1+\tau)\frac{e^{Bf(m_*)}}{Z_W}\!\left[1+\frac{\tau^2}{(1+\tau)^2}\,\eta(\overline{c})\Big(\!\big\langle f\big\rangle_{\!W}-f(m_*)\Big)\right]\!\frac{dm_*}{d\tau}\,, %
}%
\end{array}
\end{equation}
where
\begin{equation} \label{58}
\begin{array}{ccc}
\displaystyle{%
   Z_W=\int\limits_0^{m_*}\!e^{B f(m)}dm, \qquad %
   \big\langle f\big\rangle_{\!W}=\frac{1}{Z_W}\int\limits_0^{m_*}\!f(m)\hspace{0.5mm}e^{B f(m)}dm, \qquad  %
   \big\langle f^{\,2}\big\rangle_{\!W}=\frac{1}{Z_W}\int\limits_0^{m_*}\!f^{\,2}(m)\hspace{0.5mm}e^{B f(m)}dm.  %
}%
\end{array}
\end{equation}
The parameter $m_*$ is defined by the formula (\ref{48}). The
calculation shows that the entropy and heat capacity in the {\it
W}\,-phase are much higher than the values of the same quantities in
the {\it K}\,-phase, so that in Fig.\,5 where the temperature
dependences of these quantities are presented, as it was noted, on
the used scale they are close to zero. Attention is drawn to the
curious fact that when approaching zero temperature, the
contribution of the concentration wave to the heat capacity per one
particle proves to be close to unity at various average
concentrations.

At the {\it K}\,\,-{\it W} transition, the entropy turns out to be
continuous and the heat capacity undergoes a jump
\begin{equation} \label{59}
\begin{array}{ccc}
\displaystyle{%
  \frac{\Delta C}{N}\equiv\frac{C_W-C_K}{N}=\frac{2}{|\tau_i|} %
  \frac{e^{B(\tau_i)f(1)}}{Z_K(\tau_i)}\!\left[1+B(\tau_i)\Big(\!\big\langle f\big\rangle_{\!K}-f(1)\Big)\right],\quad\,\, (i=1,2). %
}%
\end{array}
\end{equation}
Thus, the transition between the {\it K}\,-\, and {\it W}\,-\,phases %
is a second-order transition. Figure 7 shows the dependence of the
magnitude of the jump in the heat capacity on the average
concentration.
\vspace{0mm} %
\begin{figure}[h!]
\vspace{-0mm}  
\includegraphics[width = 7.4cm]{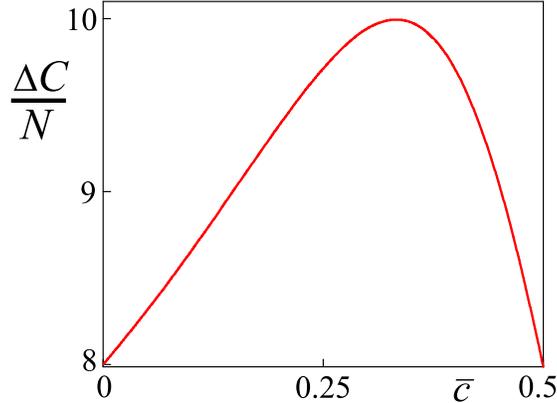} 
\vspace{-3mm} %
\caption{\label{fig07} 
Dependence of the jump in the heat capacity $\Delta C=C_W(T_1)-C_K(T_1)$ %
at the transition between the {\it K}\,-\, and {\it W}\,-\,phases on
the average concentration.
}%
\end{figure}

\noindent Note that the jump in the heat capacity remains finite
even in the limit of low concentrations, although, as mentioned
above, the condition of applicability of the model and the classical
description of the solution may be violated here.

\section{Conclusion}\vspace{-2.5mm} %
The thermodynamics of equilibrium phase transitions of the binary
solution into spatially inhomogeneous states is investigated
theoretically taking into account nonlinear effects. It is shown
that as a result of a loss of stability of the homogeneous {\it
H}\,-phase with decreasing temperature, at first there occurs the
stratification of the solution for which the concentration
distribution is described by the dependence in the form of a kink
(\ref{31}) (the {\it K}\,-phase). After reaching the maximum
possible stratification at a lower temperature there occurs a
second-order phase transition into the state with a concentration
wave which is described by the periodic solution (\ref{32}) %
(the {\it W}\,-phase). The function of distribution of states in the
inhomogeneous solution is introduced and with its help the
contribution of inhomogeneity to the entropy and heat capacity is
calculated. The phase diagram in the coordinates
temperature\,--\,average concentration is constructed. The proposed
approach can be used to describe transitions into spatially
inhomogeneous phases in other many-particle systems.

\section*{List of Notations}

\begin{tabular}{ l l }
$N$         & total number of particles          \\
$N_1, N_2 $ & numbers of particles of two types  \\
$n$         & total density of number of particles   \\
$V$         & volume of a solution               \\
$T$         & temperature                        \\
$\overline{c}=c_1$   &  average concentration of the first component       \\ %
$c=c({\bf r})$       &  local concentration of the first component         \\ %
$\delta c({\bf r})$  &  deviation of the concentration from the average value \\ %
$\varphi$     & free energy per one particle                               \\ %
$\varphi_0$   & contribution of the homogeneous state to $\varphi$         \\ %
$\overline{F}$         &  total free energy in the state of incomplete thermodynamic equilibrium \\ %
$F$                    &  total free energy in the equilibrium state                             \\ %
$K,a_2,a_3,a_4$        & coefficients in the expansion of $\varphi$ in $\big(\nabla\delta c\big)^2$ and powers of $\delta c$  \\ %
$T_p$                  & temperature at which the coefficient $a_2$ changes sign                                              \\ %
$\tau=(T-T_p)\big/T_p$ & parameter of dimensionless temperature, such that $a_2=a_0\tau$                                            \\ %
$L, A$                 & length and area of a sample      \\ %
$x$                    & one spatial coordinate on which the concentration depends  \\ %
$U(\delta c), C$       & effective field and constant of integration in the equation for $\delta c$      \\ %
$\tilde{c}=g^{-1/2}c$  & reduced concentration                               \\ %
$\tilde{\overline{c}}=g^{-1/2}\overline{c}$  & reduced average concentration \\ %
$\delta\tilde{c}=\tilde{c}-\tilde{\overline{c}}$    & reduced deviation of the concentration from the average value                               \\ %
$\tilde{x}=x/\xi_0$    & dimensionless coordinate, where $\xi_0$ is the correlation length  \\ %
$g,b,\varepsilon$      & dimensionless parameters, defined by the formula (12)              \\ %
$\tilde{U}(\delta \tilde{c}; b,\tau)$  & reduced effective field, given by the formula (15)                    \\ %
$\tilde{U}(\tilde{c}; \tilde{\overline{c}},b,\tau )$ & reduced effective field as a function of $\tilde{c}$, given by (18) in general case    \\ %
\quad                  &  and by (19) in the case $b=0$ \\ %
$\tilde{c}_{m1}, \tilde{c}_{m2}$  &  maximums of  $\tilde{U}(\tilde{c})$ for the case $b=0$ \\ %
$\tilde{c}_{01}, \tilde{c}_{02}$  &  zeros of  $\tilde{U}(\tilde{c})$  for the case $b=0$   \\ %
$V\big(|{\bf r}-{\bf r}'|\big)$   & interaction potential between particles                 \\ %
$V_0=\!\int d{\bf r}V(r)$         & constant relating to the interaction potential \\ %
$v_0, v_m, r_0$                   & constants of the potential $V(r)$, defined by the formula (25) \\ %
$\varepsilon_m$                   & maximum possible value of the parameter $\varepsilon$          \\ %
$L_K$                  & width of the transition region in the kink (31)                           \\ %
$m(\tau,\varepsilon)$  & parameter (35) entering into the elliptic functions and integrals         \\ %
$m_*(\tau)$            & $m(\tau,\varepsilon_m)$, i.e. maximum value of $m$ at given $\tau$        \\ %
$\phi(T,m)$            & ``one-particle'' distribution function over states $m$    \\ %
$f(m)$                 & characteristic function (37)                              \\ %
$E$                    & average energy contribution due to inhomogeneous states   \\ %
$S$                    & entropy contribution due to inhomogeneous states          \\ %
$C$                    & heat capacity contribution due to inhomogeneous states    \\ %
$T_1,T_2$              & temperatures of the phase transition between the inhomogeneous    \\ %
\quad                  & {\it K}\,-  \,and {\it W}\,-phases, given by the formula (52)     \\ %
$\Delta C$             &  jump of the heat capacity at the {\it K}\,\,-{\it W} transition, given by (59)      \\  %
\end{tabular}


\end{document}